\documentclass{aa}
\usepackage{graphicx}
\usepackage{txfonts}
\usepackage{natbib}
\begin{document}
\voffset=-0.5cm
\newcommand{\be}{\begin{equation}}
\newcommand{\ee}{\end{equation}}

\title{On the Relevance of Compton Scattering for the Soft X-ray
Spectra of Hot DA White Dwarfs}

\author{
V.~Suleimanov\inst{1,2,5},
J.~Madej\inst{3},
J.J.~Drake\inst{4},
T.~Rauch\inst{1} and 
K.~Werner\inst{1}}

\offprints{V.~Suleimanov}
\mail{e-mail: suleimanov@astro.uni-tuebingen.de}

\institute{
Institut f\"ur Astronomie und Astrophysik, Universit\"at T\"ubingen, Sand 1,
 72076 T\"ubingen, Germany
\and
Kazan State University, Kremlevskaja str., 18, Kazan 420008, Russia
\and
Warsaw University Observatory, Al. Ujazdowskie 4,
   00-478 Warsaw, Poland
\and
Smithsonian Astrophysical Observatory, MS-3, 60 Garden Street, Cambridge,
    MA 02138, USA
\and 
Kazan Branch of Isaac Newton Institute, Santiago, Chile
}

\date{Received xxx / Accepted xxx}

   \authorrunning{Suleimanov et al.}
   \titlerunning{Compton Scattering in Hot DA White Dwarfs}

\abstract
{}
{We re-examine the effects of Compton scattering on the
emergent spectra of hot DA white dwarfs in the soft X-ray range.
Earlier studies 
have implied that sensitive X-ray observations at
wavelengths $\lambda < 50$~\AA\ might be capable of probing the flux
deficits predicted by the redistribution of electron-scattered X-ray
photons toward longer wavelengths.}
{We adopt two independent numerical approaches to the inclusion of
Compton scattering in the computation of pure hydrogen atmospheres in 
hydrostatic equilibrium.   One employs the Kompaneets diffusion
approximation formalism,
while the other uses the cross-sections and redistribution functions
of Guilbert.  Models and emergent spectra are computed for stellar
parameters representative of HZ\,43 and Sirius~B, and for models with an
effective temperature $T_{\rm eff} = 100\, 000$ K.}
{The
differences between emergent spectra computed for Compton and Thomson
scattering cases are completely negligible  in the case of  both HZ\,43
and Sirius~B models,  and are also negligible 
for all practical purposes for models
with temperatures as high as $T_{\rm eff} = 100\, 000$ K.  
Models of the soft X-ray flux from these stars
are instead dominated by uncertainties in their fundamental
parameters.}
{}

\keywords{radiative transfer -- scattering --  methods: numerical --
 (stars:) white dwarfs -- stars: atmospheres -- X-rays: stars}

\maketitle
%

\section{Introduction}

The scattering of radiation by free electrons is one of the dominant
sources of continuous opacity in the atmospheres of hot white dwarfs.
Most model atmosphere calculations adopt the classical Thomson
isotropic scattering approach, whereby only the direction of photon
propagation changes as the result of a scattering event.  More
rigorously, finite electron mass instead implies that both momentum
and energy exchange should actually occur.

The effects of Compton scattering on white dwarf model
atmospheres was first investigated in detail by \citet{Madej:94}, who
found that pure hydrogen models with 
temperatures of $10^5$~K show a significant depression of the X-ray
continuum for wavelengths $< 50$~\AA.  Effects for models containing
significant amounts of helium, or helium and heavier elements, were
found to be much smaller or negligible, in keeping with expectations
based on the relative importance of electron scattering as an opacity
source as opposed to photoelectric absorption.

In a later paper, \citet{Madej:98} computed the effects of Compton
scattering for a model corresponding to the parameters of the DA white
dwarf HZ\,43.  Differences between Compton and Thomson scattering model
spectra were apparent for $\lambda < 50$~\AA, and grew to orders of
magnitude by 40~\AA .  While current X-ray
instrumentation is not sufficiently sensitive to study the spectra of
even the brightest DA white dwarfs in any detail at wavelengths
$\lambda < 50$~\AA, the spectral differences implied by the more
rigorous Compton redistribution formalism will be of interest to more
sensitive future missions.  Moreover, the {\it Chandra} Low Energy
Transmission Grating Spectrometer (LETG+HRC-S) effective area
calibration is based on observed spectra of HZ\,43 and Sirius~B at
wavelengths $\lambda > 60$~\AA\ \citep{Pease.etal:03}.  It is
therefore of current topical interest to re-examine the influence of
Compton scattering for these stars and determine whether any
significant differences might be discernible between Thomson and
Compton scattering in the LETGS bandpass.

In this paper, we perform two independent and rigorous tests of the
influence of Compton Scattering on the emergent spectra of hot DA
white dwarfs.  Our methods of calculation are outlined in
\S\ref{s:methods}, while results and conclusions are briefly discussed
in  \S\S\ref{s:results} - \ref{s:conclusions}.

\section{Computational Methods}
\label{s:methods}

In both our numerical approaches, outlined below, we computed model
atmospheres of hot white dwarfs subject to the constraints 
of hydrostatic and radiative equilibrium assuming planar geometry
using standard methods \citep[e.g.\ ][]{Mihalas:78}.
The equation of state of an ideal gas used assumes local thermodynamic
equilibrium (LTE), and therefore did not include terms describing the
local radiation field.

The model atmosphere structure for a hot WD is described by the 
hydrostatic equilibrium equation,
\be \label{e:hyd}
  \frac {d P_{\rm g}}{dm} = \frac{GM_{\rm wd}}{R^2_{\rm wd}} -
4 \pi \int_0^{\infty}  H_{\nu} \, \frac{k_{\nu}+\sigma_\nu}{c} \, d\nu,
\ee
where $k_{\nu}$ is opacity per unit mass due to free-free, bound-free and 
bound-bound
transitions, $\sigma_\nu$ is the electron (Compton) opacity, $H_{\nu}$ is
Eddington flux, $P_{\rm g}$ is a gas pressure, and $m$ is column density
\be
     dm = -\rho \, dz \, .
\ee
Variable $\rho$ denotes the gas density and $z$ is the vertical distance.
As is obvious from Eqn.~\ref{e:hyd}, the structure of the atmosphere is
coupled to the radiation field and the structure and radiative
transfer equations need to be solved simultaneously under the
constraint of radiative equilibrium.
 
In the Thomson approximation, in which no energy or momentum between
photons and electrons is exchanged, $\sigma_\nu=\sigma_{\rm e}$, where
$\sigma_{\rm e}$ is the classical Thomson opacity.

\subsection{Method 1}

In our first approach, Compton scattering is taken into account in 
the radiation transfer equation using the Kompaneets operator 
\citep{Kompaneets:57, Zavlin.Shibanov:91, Grebenev.Sunyaev:02}:
\begin{eqnarray} \label{rtr}
   \frac{\partial^2 f_{\nu} J_{\nu}}{\partial \tau_{\nu}^2} =
\frac{k_{\nu}}{k_{\nu}+\sigma_{\rm e}} \left(J_{\nu} - B_{\nu}\right) -
 \frac{\sigma_{\rm e}}{k_{\nu}+\sigma_{\rm e}} \frac{kT}{m_{\rm e} c^2}
\times \\ \nonumber
 x
\frac{\partial}{\partial x} \left(x \frac{\partial J_{\nu}}{\partial x} -
3J_{\nu} + \frac{T_{\rm eff}}{T} x J_{\nu} \left( 1 + \frac{CJ_{\nu}}{x^3}
\right) \right),
\end{eqnarray}
where $x=h \nu /kT_{\rm eff}$ is the dimensionless frequency,
$f_{\nu}(\tau_{\nu}) \approx 1/3$ is the variable Eddington factor, $J_{\nu}$
is the mean intensity of radiation, $B_{\nu}$ is the black body (Planck)
intensity,
$T$ is the local
electron temperature, $T_{\rm eff}$ is the effective temperature of WD,
 and $C=c^2 h^2~/~2(kT_{\rm eff})^3$. The optical
depth $\tau_{\nu}$ is defined as
\be
    d \tau_{\nu} = (k_{\nu}+\sigma_{\rm e}) \, dm.
\ee
These equations have to be completed by the energy balance equation
\begin{eqnarray}  \label{econs}
 \int_0^{\infty} k_{\nu}\left(J_{\nu} - B_{\nu}\right) d\nu -
 \sigma_{\rm e} \frac{kT}{m_{\rm e} c^2} \times \\ \nonumber
\left( 4 \int_0^{\infty} J_{\nu} \,
d\nu - \frac{T_{\rm eff}}{T} \int_0^{\infty} x J_{\nu}
(1+\frac{CJ_{\nu}}{x^3}) \, d\nu \right)=0,
\end{eqnarray}
the ideal gas law
\be   \label{gstat}
    P_{\rm g} = N_{\rm tot} kT,
\ee
where $N_{\rm tot}$ is the number density of all particles, and also
by the particle and charge conservation equations.  We assume local
thermodynamical equilibrium (LTE) in our calculations, so the number
densities of all ionisation and excitation states of all elements have
been calculated using Boltzmann and Saha equations.

For solving the above equations and computing the model atmosphere we
used a version of the computer code ATLAS \citep{Kurucz:70,Kurucz:93},
modified to deal with high temperatures; see \citet{Ibragimov.etal:03}
and \citet{Swartz.etal:02} for further details.  This code was also
modified to account for Compton scattering.

The scheme of calculations is as follows.  First of all, the input
parameters of the WD are defined: the effective temperature $T_{\rm
eff}$ and surface gravity $g=GM_{\rm wd}/R^2_{\rm wd}$.  Then a
starting model using a grey temperature distribution is calculated.
The calculations are performed with a set of 98 depth points $m_{\rm
i}$ distributed logarithmically in equal steps from $m\approx
10^{-6}$~g~cm$^{-2}$ to $m_{\rm max}$. The appropriate value of
$m_{\rm max}$ is found from the condition $\sqrt{\tau_{\nu,\rm
b-f,f-f}(m_{\rm max})\tau_{\nu}(m_{\rm max})} >$ 1 at all frequencies.
Satisfying this equation is necessary for the inner boundary condition
of the radiation transfer.

For the starting model, all number densities and opacities at all
depth points and all frequencies (we use 300 logarithmically
equidistant frequency points) are calculated. The radiation transfer
equation (\ref{rtr}) is non-linear and is solved iteratively by the
Feautrier method \citep[][see also \citealt{Zavlin.Shibanov:91,
Pavlov.etal:91,Grebenev.Sunyaev:02}]{Mihalas:78}.  We use the last
term of the equation (\ref{rtr}) in the form
$xJ_{\nu}^i(1+CJ_{\nu}^{i-1}/x^3)$, where $J_{\nu}^{i-1}$ is the mean
intensity from the previous iteration.  During the first iteration we
take $J_{\nu}^{i-1}=0$.  Between iterations we calculate the variable
Eddington factors $f_{\nu}$ and $h_{\nu}$, using the formal solution
of the radiation transfer equation in three angles at each frequency.
Usually 2-3 iterations are sufficient to achieve convergence.

We used the usual condition at the outer boundary
\be
    \frac{\partial J_{\nu}}{\partial \tau_{\nu}} = h_{\nu} J_{\nu},
\ee
where $h_{\nu} \approx 1/2$ is surface variable Eddington factor.
The inner boundary condition is
\be
   \frac{\partial J_{\nu}}{\partial \tau_{\nu}} =
 \frac{\partial B_{\nu}}{\partial \tau_{\nu}}.
\ee
The outer boundary condition is found from the lack of incoming
radiation at the WD surface, and the inner boundary condition is obtained
from the diffusion approximation $J_{\nu} \approx B_{\nu}$ and $H_{\nu}
\approx 1/3 \times \partial B_{\nu}/\partial \tau_{\nu}$.

The boundary conditions along the frequency axis are
\be  \label{lbc}
      J_{\nu} = B_{\nu}
\ee
at the lower frequency boundary ($\nu_{\rm min}=10^{12}$ Hz,
 $h\nu_{min}$ $\ll kT_{\rm eff}$) and
\be  \label{hbc}
x \frac{\partial J_{\nu}}{\partial x} - 3J_{\nu} + \frac{T_{\rm eff}}{T} x
J_{\nu} \left( 1 + \frac{CJ_{\nu}}{x^3} \right)=0
\ee
at the higher frequency boundary ($\nu_{\rm max}\approx 3 \cdot
10^{17}$ Hz, $h\nu_{\rm max} \gg kT_{\rm eff}$).  Condition
(\ref{lbc}) means that at the lowest energies the true opacity
dominates the scattering $k_{\nu} \gg \sigma_{\rm e}$, and therefore
$J_{\nu} \approx B_{\nu}$. Condition (\ref{hbc}) means that there is
no photon flux along frequency axis at the highest energy.

The solution of the radiative transfer equation (\ref{rtr}) was checked
for the energy balance equation (\ref{econs}) together with the surface
flux condition
\be
    4 \pi \int_0^{\infty} H_{\nu} (m=0) d\nu = \sigma T_{\rm eff}^4 = 4 \pi
   H_0.
\ee
The relative flux error as a function of depth,
\be
     \varepsilon_{H}(m) = 1 - \frac{H_0}{\int_0^{\infty} H_{\nu} (m) d\nu},
\ee
was calculated, where $H_{\nu} (m)$ is radiation flux at given depth.
This latter quantity was found from
the first moment of the radiation transfer equation:
\be
    \frac{\partial f_{\nu} J_{\nu}}{\partial \tau_{\nu}} = H_{\nu}.
\ee
Temperature corrections were then evaluated using three different
procedures.  The first is the integral $\Lambda$-iteration method,
modified for Compton scattering, based on the energy balance equation
(\ref{econs}).  It is valid in the upper atmospheric layers.  The
second one is the Avrett-Krook flux correction, which uses the
relative flux error, and is valid in the deep layers.  And the third
one is the surface correction, which is based on the emergent flux
error.  See \citet{Kurucz:70} for a detailed description of the
methods.

The iteration procedure is repeated until the relative flux error is
smaller than 1\%, and the relative flux derivative error is smaller
than 0.01\%. As a result of these calculations, we obtain the
self-consistent WD model atmosphere together with the emergent
spectrum of radiation.

\begin{figure}
\includegraphics[width=1.0\columnwidth]{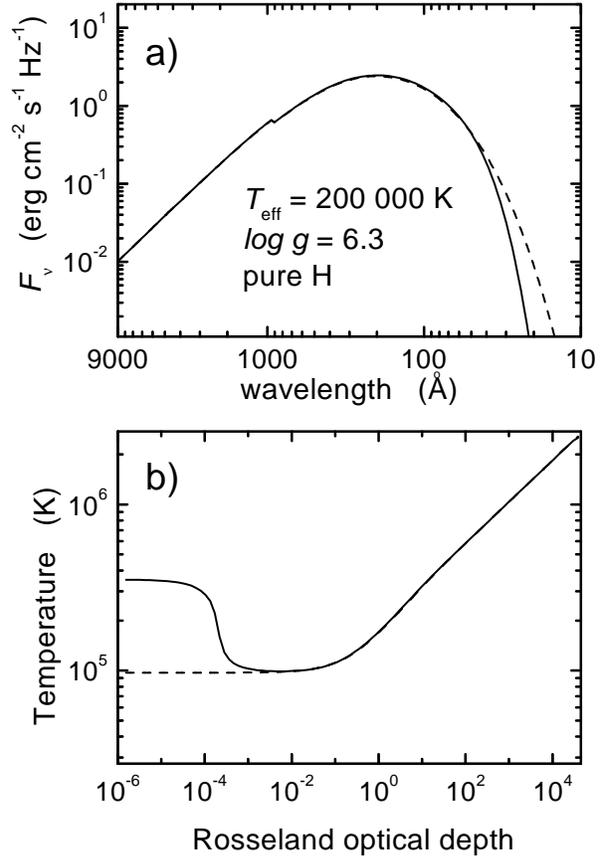}
\caption{\label{f:sul200} 
Spectra (a) and temperature structures (b) of a hot DA white dwarf model 
atmosphere with $T_{\rm eff}= 200\, 000$~K and $\log~ g=6.3$ with (solid 
line) and without (dashed line) Compton scattering taken into 
account.} 
\end{figure}

Our method of calculation was tested on a model of bursting neutron
star atmospheres \citep{Pavlov.etal:91,Madej:91}, and a model DA white
dwarf atmosphere with $T_{\rm eff}=2\cdot 10^5$ K, $\log~g = 6.3$
\citep{Madej:94}.  Agreement with the earlier calculations is
extremely good.  We show the emergent spectrum from the
latter calculation in Fig.\ref{f:sul200}.

\subsection{Method 2} 

Our second approach adopts the equation of transfer for absorption
and scattering presented by \citet{Sampson:59} and \citet[][see
  Eqn. (2.167)]{Pomraning:73}.
The equation of transfer can be expressed in the form
\begin{eqnarray}
{\partial ^2 f_\nu J_\nu \over {\partial \tau_\nu^2 }} & = &
 {k_\nu \over {k_\nu + \sigma_\nu} } \, (J_\nu - B_\nu) + \nonumber  \\
 & + & {\sigma_\nu \over {k_\nu + \sigma_\nu} } \, J_\nu \,
\int \limits _{0}^{\infty}  \Phi (\nu , \nu ^\prime )
\, \left(1+ {c^2 \over {2h{\nu ^\prime }^3 }} J_{\nu ^\prime} \right)\,
d\nu ^\prime  \nonumber \\
 & - & {\sigma_\nu \over {k_\nu + \sigma_\nu} } \, 
 \left( 1+ {c^2 \over {2h\nu ^3}} J_\nu \right)\, \times \nonumber \\
 & & \int \limits _{0}^{\infty} \Phi (\nu ,\nu ^\prime) J_{\nu ^\prime}
\, {\left( {\nu \over {\nu ^\prime }} \right) } ^3
\exp \left[ -{{h(\nu - {\nu ^\prime }) }\over {kT}} \right]
\, d\nu ^\prime  .
\label{equ:tau}
\end{eqnarray}
This transfer equation is written on the monochromatic optical depth 
scale $d\tau_\nu=-(k_\nu+\sigma_\nu) \, \rho \, dz$.
The variable $J_\nu$ denotes the energy-dependent mean intensity of
radiation. The function $\Phi(\nu,\nu ^\prime)$ denotes the zeroth angular
moment of the angle-dependent differential cross-section, normalized to unity
(see below). 

Transformation of the equation of transfer by \citet{Pomraning:73} to
the Eq.~\ref{equ:tau} and the definition of required angular approximations 
for Compton scattering in a stellar atmosphere was outlined by 
\citet{Madej:91,Madej:94} and \citet{Madej.etal:04}. 

Our equations and theoretical models of Method 2 use detailed differential
cross-sections for Compton scattering, $\sigma(\nu \to \nu^\prime,
\vec{n} \cdot \vec{n^\prime })$, which were taken from \citet{Guilbert:81}. 
Cross-sections correspond to scattering in a gas of free electrons with 
relativistic thermal velocities, and they are also completely valid at low 
temperatures.
Differential cross-sections were then integrated numerically to obtain large
grids of Compton scattering opacity coefficients $\sigma_\nu$
\begin{equation}
\sigma_\nu = \oint_{\omega^{\prime}}
\frac{d\omega^\prime}{4\pi} \int^{\infty}_{0} d\nu^{\prime} \,
\sigma(\nu \to \nu^\prime, \vec{n} \cdot \vec{n^\prime }) \, ,
\end{equation}
and grids of angle-averaged Compton scattering redistribution
functions $\Phi (\nu, \nu^\prime)$
\begin{equation}
\Phi(\nu,\nu^\prime) = {1 \over {\sigma_\nu}}
  \oint_{\omega^{\prime}} \frac{d\omega^\prime}{4\pi} \,
  \sigma(\nu \to \nu^\prime, \vec{n} \cdot \vec{n^\prime })  \, .
\end{equation}  

The scattering frequency redistribution function was introduced by 
\citet{Pomraning:73}---see Eqn. (7.95) of his book---and it represents
the normalized probability density of scattering from a given frequency 
$\nu$ to the outgoing frequency $\nu^\prime$.

Note that Eq.~\ref{equ:tau} also includes  stimulated  scattering terms,
$1+(c^2/2h \nu^3) \, J_\nu$, which ensure the physically correct
description of Compton scattering.  The actual equations and
calculations used here strongly differ from those in \citet{Madej:98}.
The physics of Compton scattering used here is also 
fundamentally different to Method 1. Note, that the latter method and that
of \citet{Madej:98}
employ the well-known Kompaneets diffusion approximation to Compton scattering
kernels. The apparent difference between Methods 1 and 2 is that they
use either the differential Kompaneets kernel (Method 1) or kernels given
by integrals over the detailed Compton scattering profiles (Method 2).

The equation of radiative equilibrium (the energy balance equation) requires
that
\begin{equation}
\int^\infty _0  H_\nu \, d\nu = \frac{\sigma_R T_{\rm eff}^4}{4\pi} \, .
\label{equ:rad0}
\end{equation}
The above condition is fulfilled in a hot stellar atmosphere, where energy 
transport by convective motions can be neglected. 

Computing derivatives of both sides of Eqn.~\ref{equ:rad0} and using the
equation of transfer, Eqn.~\ref{equ:tau}, one can obtain the alternative 
energy balance equation
\begin{eqnarray}
 \int\limits_{0}^{\infty} k_\nu \, (J_\nu\, - B_\nu) \,  d\nu +
  \int\limits_{0}^{\infty} \sigma_\nu J_\nu d\nu \int\limits_{0}^{\infty}
  \Phi (\nu,\nu^\prime ) \, \left(1+ {c^2 \over {2h{\nu ^\prime }^3 }}
  J_{\nu^\prime} \right)\, & d\nu^\prime\,  \nonumber \\
 \hskip-70mm - \int\limits_{0}^{\infty} \sigma_\nu \left( 1+ {c^2 \over
  {2h\nu ^3}} J_\nu \right)\, d\nu & \nonumber \\ 
   \int\limits_{0}^{\infty} J_{\nu^\prime}
  \, \Phi(\nu,\nu^\prime) \, {\left( {\nu \over {\nu^\prime}} \right) }^3
  \exp \left[ -{{h(\nu - {\nu^\prime }) }\over {kT}} \right]
\, d\nu^\prime = 0\, ,&  
\label{equ:rad}
\end{eqnarray}
which can be compared with its analog, the Eqn.~\ref{econs} of Method 1.

The computer code {\sc atm21} used for the model calculations was
described in detail in \citet{Madej.Rozanska:00b} and
\citet{Madej.etal:04}.  The structure of the code is based on the
partial linearization scheme by \citet{Mihalas:78}, in which
corrections of temperature $\Delta T$ and the function $\Delta \Phi$
are built into the equation of transfer.  The high numerical accuracy
and very good convergence properties of the {\sc atm21} code are vital
for the present paper, and allowed us to compute accurate spectra for
atmospheric parameters appropriate for the white dwarfs HZ\,43 and
Sirius B using Method 2, outlined above.  These calculations supersede
less accurate X-ray spectra of HZ\,43 which were presented in the
earlier paper \citep{Madej:98}.

For the present research, the {\sc atm21} code included 
numerous bound-free LTE opacities of neutral hydrogen and free-free
opacity of ionized hydrogen, which always remains in LTE. 
No hydrogen lines were included in the actual computations. In each
temperature iteration the code solves the equation of hydrostatic
equilibrium to obtain stratifications of gas pressure $P_g$ and density 
$\rho$ in the model atmosphere. After that the {\sc atm21} code solves
the set of coupled equations of  radiative transfer with implicit 
temperature corrections and finds the stratification of $\Delta T$ in
the model atmosphere. The equation of radiative transfer was solved
using the
Feautrier method and the technique of variable Eddington factors
\citep{Mihalas:78}.  Boundary conditions along the $\tau_\nu$ axis were 
the same as in Method 1. 
Explicit expressions for the temperature corrections, $\Delta T$, can be
found, e.g.,\ in \citet{Madej.etal:04}.

\section{Computations and Results}
\label{s:results}

Using our two independent methods, we investigated the effects of
Compton scattering on the emergent spectra of three pure H white
dwarf models.  These were: models appropriate for the well-known DA
white dwarfs Sirius~B (assuming 
$T_{\rm eff}=24\,700$~K,  $\log~g=8.6$) and HZ\,43 ($T_{\rm
  eff}=50\,000$~K, $\log~g=8.0$), together with a significantly hotter  
model with $T_{\rm eff}=1\cdot 10^5$ K and $\log~g = 6$ and $8$.  These
models cover the effective temperature range relevant to pure H DA 
atmospheres and specifically address the question of whether Compton
scattering might be relevant to the 
X-ray bright DA white dwarfs Sirius~B and HZ\,43 that are central to
the low-energy calibration of {\it Chandra} \citep{Pease.etal:03}.

\begin{figure}
\includegraphics[width=1.0\columnwidth]{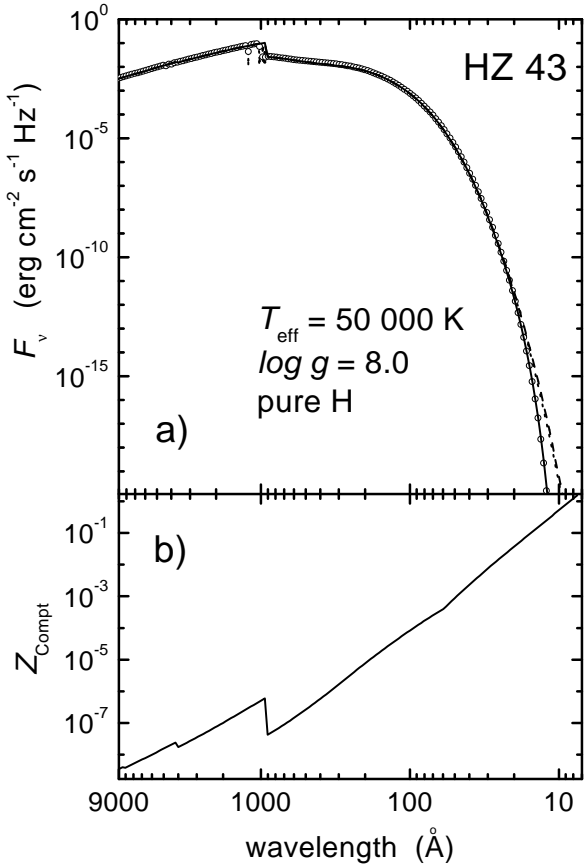}
\caption{\label{f:sulhz43} 
Spectra (a) of the DA white dwarf HZ\,43 model atmosphere with (solid line) 
and without (dashed line) Compton scattering  using Method 1.  Also shown are the spectra 
of the non-LTE model atmosphere (dotted line) and  the spectra of non-LTE 
model atmosphere with Compton scattering (open circles). Run (b) of 
Comptonisation parameter $Z_{\rm Compt}$ with wavelength (see
Eqn.~\ref{e:ycomp}).
}
\end{figure}

\begin{figure}
\includegraphics[width=1.0\columnwidth]{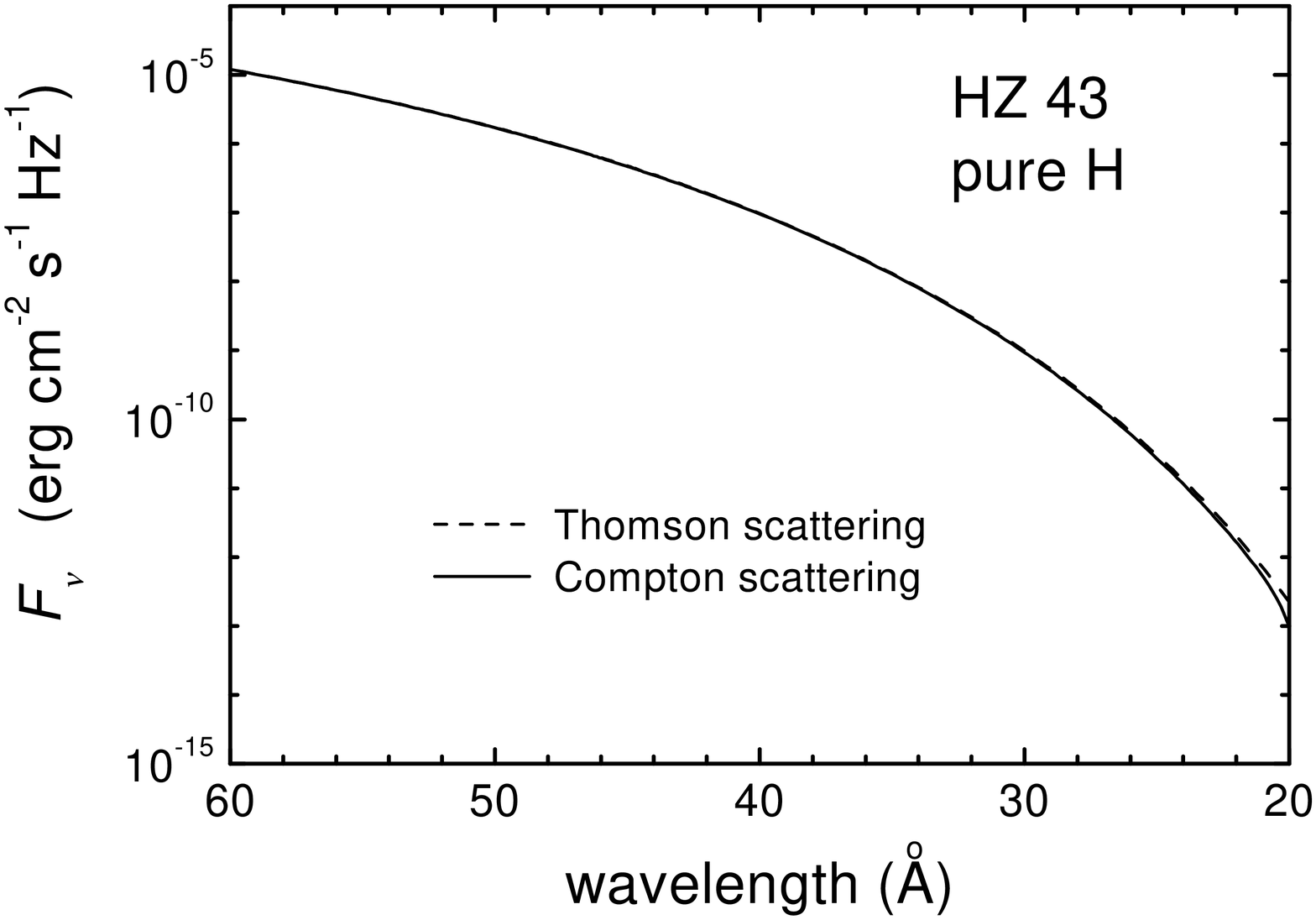}
\caption{Emergent X-ray spectra for a pure hydrogen model atmosphere
appropriate for the parameters of HZ\,43 computed for classical Thomson
scattering (dashed) and Compton scattering (solid) using Method 2 (the
method of \citealt{Madej.etal:04}).
\label{f:madhz43}
}
\end{figure}

\begin{figure}
\includegraphics[width=1.0\columnwidth]{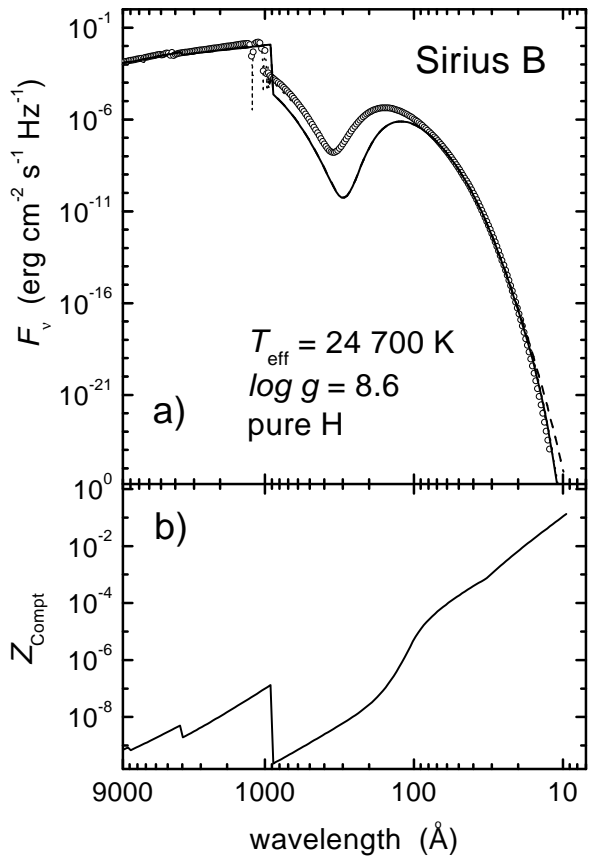}
\caption{\label{f:sulsb} 
The same as Fig.~\ref{f:sulhz43} but here for the DA white dwarf Sirius~B
LTE model atmosphere. The dotted line lies everywhere under the open circules
(except for hydrogen lines).} 
\end{figure}

\begin{figure}
\includegraphics[width=1.0\columnwidth]{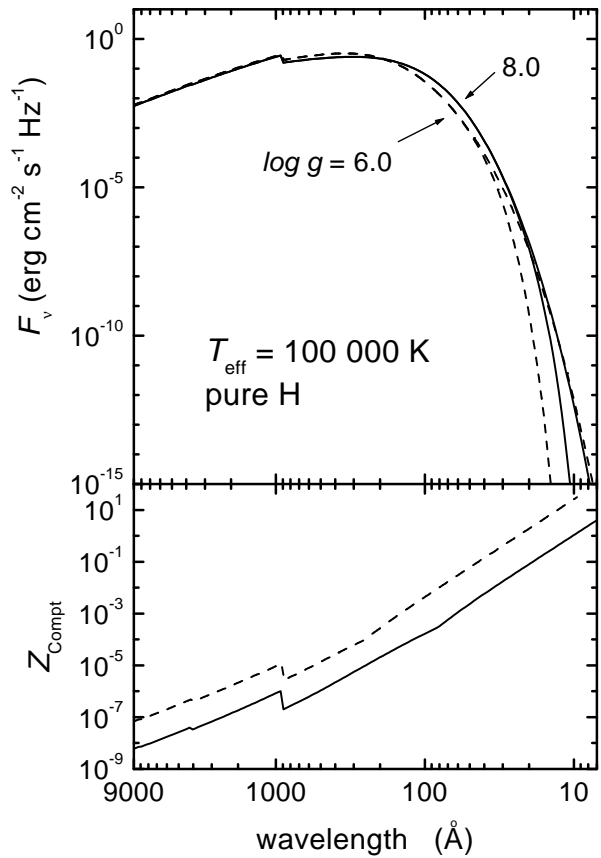}
\caption{\label{f:sulf4} 
Spectra (a) of the DA white dwarf $T_{\rm eff} = 100\,000$ K model 
atmospheres with  
and without Compton scattering. Solid lines correspond to model with
$\log~g=8.0$, and dashed lines correspond to model with $\log~g = 6.0$. 
Softer high energy tails correspond to models with Compton effect. Run (b) of 
Comptonisation parameter  $Z_{\rm Compt}$ with wavelength.} 
\end{figure}

The results of our calculations from both methods are presented in
Figures~\ref{f:sulhz43}-\ref{f:sulf4}.  In Fig.~\ref{f:sulhz43}a the
spectra of the model atmosphere for the DA white dwarf HZ\,43 ($T_{\rm
eff}=50\,000$~K, $\log~g=8.0$) computed using Method~1 with (solid
line) and without (dashed line) Compton effects are shown.  We also 
calculated a non-LTE model atmosphere for HZ\,43 using the T\"ubingen Model
Atmosphere Package ({\sc
TMAP}) \citep{Werner.etal:03} and computed  the radiation transfer
equation (\ref{rtr}) with Compton scattering  using this non-LTE model
atmosphere structure.  The corresponding spectra are shown in
Fig.~\ref{f:sulhz43}a by the dotted line and by open circles.

Calculations using Method~2 in LTE are illustrated in
Fig.~\ref{f:madhz43}.  The differences between Compton and Thomson
models in these calculations are very similar to the differences found
in Method~1 in Fig.~\ref{f:sulhz43}a, and are significant only at
wavelengths $< 25$~\AA.
The flux in the
models with Compton scattering (Method 1) is smaller than the flux in the model
without Compton scattering by 0.5\% at 50~\AA, 1.3\% at 40~\AA, 7\%
at 30~\AA, and 40\% at 20~\AA.  Similarly, model computations performed with
Method 2 
yield the following differences for HZ\,43: 
0.5\% at 50~\AA, 1.4~\% at 40~\AA, 5\% at 30~\AA, and 56\% at 20~\AA. 

These results can be understood more clearly if we
consider the Comptonisation parameter $Z_{\rm Compt}$: 
\be Z_{\rm
Compt} = \frac{h\nu}{m_{\rm e}c^2} \max((\tau_{\rm e}^*)^2,\tau_{\rm
e}^*), 
\label{e:ycomp}
\ee 
where $h\nu/m_{\rm e}c^2$ is the relative photon energy lost
during one scattering event off a cool electron, $\max((\tau_{\rm
e}^*)^2,\tau_{\rm e}^*)$ is the number of scattering events the photon
undergoes before escaping, $\tau_{\rm e}^*$ is the Thomson optical depth,
corresponding to the depth where escaping photons of a given frequency
are created. The Comptonisation parameter is, then, a representation
of the influence of Compton down-scattering on the emergent spectrum:
significant Compton effects are expected if the Comptonisation
parameter approaches unity \citep{Rybicki.Lightman:79}. In
Fig.\ref{f:sulhz43}b the dependence of $Z_{\rm Compt}$ on the
wavelength is shown.  It is clear that the Comptonisation parameter is
very small down to 25 \AA, and this is indeed reflected in the emergent
spectrum.

Similar results were obtained for the DA white dwarf Sirius~B
(assuming $T_{\rm eff}=24\,700$~K, $\log~g=8.6$).  The corresponding
spectra and the Comptonisation parameter are shown in
Fig.~\ref{f:sulsb}. The effective temperature of this star is smaller,
and the surface gravity is larger, therefore, Compton scattering is
even less significant than for HZ\,43.  We also calculated a non-LTE
model atmosphere for Sirius~B using the {\sc TMAP} and computed
 the radiation transfer equation (\ref{rtr}) with Compton scattering
using this non-LTE model atmosphere structure, as for HZ\,43.  The
corresponding spectra are shown in Fig.~\ref{f:sulsb}a by the dotted
line and by open circles.  Compton scattering changes only very
slightly the emergent spectrum at wavelengths below 20 \AA.

In the Fig.~\ref{f:sulf4} we present the spectra of the hot DA white
dwarfs model atmospheres ($T_{\rm eff} = 100\,000$ K, $\log~g = 6$ and
$8$) with and without Compton scattering.  It is obvious that Compton
effects are more significant for these hot DA models (especially with
low surface gravity) than for Sirius~B and HZ\,43, but visible effects
still do not occur for wavelengths $>50$~\AA. Of course, $\log~g=6$ is
not a so realistic value for white dwarfs, but we calculate this model to
 demonstrate the dependence of the Compton effect on the surface gravity.

\section{Discussion}

The differences we find here between emergent spectra computed using
Thomson and Compton scattering are essentially negligible for practical
purposes.  Such differences are in fact much smaller than the
predicted spectral uncertainties resulting from uncertainties in the
current knowledge of the fundamental parameters of DA white
dwarfs - even the best known examples such as HZ\,43 and Sirius~B. 
The spectrum in this Wien tail region is especially sensitive to
uncertainties in the effective temperature.  

Comparison of the calculations presented here using both Methods 1 and
2 with the earlier work of \citet{Madej:98} do, however, reveal some
significant differences.  For HZ\,43, Fig.~3 of \citet{Madej:98}
suggests a large X-ray flux deficit due to Compton scattering of
photons to longer wavelengths, with a precipitous decline in emergent
flux at $\sim 40$~\AA.  However, our Figs.~\ref{f:sulhz43} and 
\ref{f:madhz43} illustrate much smaller effects. 

A careful examination of the \citet{Madej:98} computer code performed
by one of us (JM) has shown that the earlier code (which used
Kompaneets scattering terms) could strongly exaggerate effects of
Compton scattering in cases when they were of only marginal
significance. This is the case for the X-ray spectrum of HZ\,43.  The
effect was caused by an approximation adopted in the solution of the
transfer equation that was quite valid for the study of X-ray burst
sources, for which the code was primarily developed, but which became 
marginally inaccurate for the case of hot DA white dwarf atmospheres.
This problem has been solved by the very stable algorithm of the new code (see
Method 2), described in \citet{Madej.Rozanska:00b}.  

One should note that the differences between this work and that of
\citet{Madej:98} are not related to the numerical approaches adopted
for Compton scattering.  Both the Kompaneets diffusion approximation
(Method 1) and the Compton scattering terms of the integral form
(Method 2) satisfactorily describe effects of Compton scattering in
X-ray spectra of hot white dwarf stars.  The results of the work
presented here using both methods supersede those of \citet{Madej:98}.

At the higher effective temperatures represented by the $T_{\rm eff} =
100\,000$ K models, the emergent spectra for Thomson and Compton
scattering begin to diverge at $\sim 50$~\AA\ for the $\log g=6$
model, and the effects are much larger toward shorter wavelengths than
for the higher gravity $\log g=8$ case.  It is of course questionable
as to whether any white dwarfs with pure H atmospheres exist with such
high effective temperatures, since in hotter stars radiative
levitation tends to enrich the atmosphere with metals.  In atmospheres
with significant metal abundances the electron scattering opacity is
insignificant compared with that due to metals, and Compton
redistribution effects are rendered irrelevant.  

Our calculations show that even in the hottest pure H atmospheres it
is highly unlikely that future X-ray observations will be sufficiently
sensitive to discern the Compton redistribution effects. Again,
uncertainties in model parameters such as effective temperature and
surface gravity, and abundances of He and trace elements, together
with uncertainties in parameters entering into the modeling calculations
themselves will dominate.  Even for models normalised to the same flux
at UV wavelengths a 1\%\ error in $T_{\rm eff}$ will induces a 20\%\
flux error at 75\AA.

\section{Conclusions}
\label{s:conclusions}

New calculations using two independent, rigorous numerical methods
confirm that the effects of Compton energy redistribution in
photon-electron scattering events are completely negligible for the
interpretation of X-ray spectra of DA white dwarfs such as Sirius B
and HZ\,43.  Differences between emergent spectra of Compton and Thomson
cases are in fact much smaller than the predicted spectral
uncertainties resulting from uncertainties in the current knowledge of
the fundamental parameters of stars such as HZ\,43.  We have found that
differences between Compton effects predicted here for HZ\,43 and the
calculations of \citet{Madej:98} are caused by approximations used in
the solution of the transfer equation in the former work; the results
presented here supersede the earlier ones.  We conclude that
current non-LTE model atmosphere spectra of hot DA white dwarfs
neglecting Compton scattering can be safely used for the
calibration low energy detectors of X-ray observatories for the
foreseeable future.

\begin{acknowledgements}

VS thanks DFG for financial support (grant We 1312/35-1) and Russian FBR
(grant 05-02-17744) for partial support of this investigation.
JM acknowledges support by the Polish Committee for
Scientific Research grant No. 1 P03D 001 26.
JJD was supported by NASA contract NAS8-39073 to the
{\em Chandra X-ray Center} during the course of this research. TR is supported by DLR (grant 50 OR 0201).

\end{acknowledgements}


\bibliographystyle{aa}
\bibliography{compton,jjdrake}

\end{document}